\documentclass[a4paper,amsmath,amssymb,prb,preprintnumbers,superscriptaddress,twocolumn,showpacs,longbibliography]{revtex4-1}

\usepackage{graphicx}

\usepackage{dcolumn}  
\usepackage{bm}         
\usepackage{amssymb}
\usepackage{amsmath}
\usepackage{epsfig}
\usepackage{xcolor}

\usepackage{xspace}

\newif\ifcom
\newif\ifdel

\comtrue                    %
\deltrue                      %

\begin{document}

\title{NanoSQUIDs from YBa$_2$Cu$_3$O$_7$/SrTiO$_3$ superlattices with bicrystal grain boundary Josephson junctions}

\author{Jianxin Lin}
\affiliation{%
Physikalisches Institut, Center for Quantum Science (CQ) and LISA$^+$,
University of T\"ubingen,
Auf der Morgenstelle 14,
72076 T\"ubingen, Germany}

\author{Julian Linek}
\affiliation{%
Physikalisches Institut, Center for Quantum Science (CQ) and LISA$^+$,
University of T\"ubingen,
Auf der Morgenstelle 14,
72076 T\"ubingen, Germany}

\author{Reinhold Kleiner}
\affiliation{%
Physikalisches Institut, Center for Quantum Science (CQ) and LISA$^+$,
University of T\"ubingen,
Auf der Morgenstelle 14,
72076 T\"ubingen, Germany}

\author{Dieter Koelle}
\email{koelle@uni-tuebingen.de}
\affiliation{%
Physikalisches Institut, Center for Quantum Science (CQ) and LISA$^+$,
University of T\"ubingen,
Auf der Morgenstelle 14,
72076 T\"ubingen, Germany}

\date{\today}

\begin{abstract} 
We report on the fabrication and characterization of nanopatterned dc SQUIDs with grain boundary Josephson junctions based on heteroepitaxially grown YBa$_2$Cu$_3$O$_7$ (YBCO)/ SiTrO$_3$ (STO) superlattices on STO bicrystal substrates.
Nanopatterning is performed by Ga focused-ion-beam milling.
The electric transport properties and thermal white flux noise of superlattice nanoSQUIDs are comparable to single layer YBCO devices on STO bicrystals.
However, we find that the superlattice nanoSQUIDs have more than an order of magnitude smaller low-frequency excess flux noise, with root-mean-square spectral density $S_\Phi^{1/2}\sim 5-6\,\mu\Phi_0/\sqrt{\rm Hz}$ at 1\,Hz ($\Phi_0$ is the magnetic flux quantum).
We attribute this improvement to an improved microstructure at the grain boundaries forming the Josephson junctions in our YBCO nanoSQUDs.
\end{abstract} 



\maketitle

\section{Introduction}
\label{sec:Introduction}

Strongly miniaturized direct current (dc) superconducting quantum interference devices (SQUIDs) with dimensions in the submicrometer range (nanoSQUIDs) are promising devices for the sensitive detection and investigation of small spin systems, as they can be used for direct detection of the magnetization reversal of individual magnetic nanoparticles (MNPs), nanotubes or nanowires \cite{Wernsdorfer01, Jamet01, Wernsdorfer09, Hao11, Buchter13, Granata16, Martinez-Perez17a, Martinez-Perez17, Martinez-Perez18, Martinez-Perez20} and for high-resolution scanning SQUID microscopy \cite{Kirtley95, Kirtley09, Vasyukov13, Koelle13, Hazra13, Kirtley16, Reith17, Wyss19, Anahory20, Uri20}.
The vast majority of nanoSQUIDs developed during the last decade are based on metallic superconductors, for which their operation temperature is limited by their transition temperature $T_{\rm c}$ to below $\sim 10\,$K.
Furthermore, metallic low-$T_{\rm c}$ superconductors typically have upper critical fields $B_{\rm c2}$ below 1\,T, which also limits the magnetic field range in which nanoSQUIDs based on them can be operated \cite{Vasyukov13, Woelbing13}.
As an alternative, the high-$T_{\rm c}$ cuprate superconductor YBa$_2$Cu$_3$O$_7$ (YBCO), with $T_{\rm c}\sim 92\,$K and $B_{\rm c2}$ well above 10\,T offers the use of YBCO nanoSQUIDs for applications within a much wider range of temperature $T$ and magnetic field $B$, as compared to nanoSQUIDs based on conventional metallic superconductors.

The fabrication of Josephson junctions (JJs) and SQUIDs based on cuprate superconductors is demanding, in particular due to the very small superconducting coherence length on the nanometer scale and concomitant high sensitivity of those materials to defects on the atomic scale.
YBCO micro- and nanoSQUIDs are typically based on single layers of epitaxially grown films, that are eventually covered with a metallic layer (e.g.~Au) for resistive shunting and protection during nanopatterning \cite{Martinez-Perez17a}.
Josephson junctions in such single layer devices are based on constrictions (cJJs) with widths down to $\sim 50\,$nm \,\cite{Arpaia14}, that yield a root-mean-square (rms) spectral density of flux noise $S_\Phi^{1/2}$ in the thermal white noise limit down to $< 450\,{\rm n}\Phi_0/{\rm Hz}^{1/2}$ at $T=18\,$K ($\Phi_0$ is the magnetic flux quantum) \cite{Arpaia17}.
Another recently developed JJ type is based on creating Josephson barriers in YBCO thin films via focused ion beam (FIB) irradiation with a He ion beam\cite{Cybart15}.
Sub-micron wide He-FIB JJs \cite{Cho18} and nanoSQUIDs \cite{Li20} have been realized, and a rms flux noise in the thermal white noise limit down to $<500\,{\rm n}\Phi_0/{\rm Hz}^{1/2}$ has been achieved at $T=4.2\,$K \,\cite{Mueller19}.
Furthermore, grain boundary (GB) Josephson junctions (GBJJs), that are formed by epitaxial growth of YBCO on a bicrystal substrate \cite{Hilgenkamp02}, provide a well established approach for realizing YBCO nanoSQUIDs in single layer devices.
The use of Ga FIB milling enables one to fabricate GBJJs down to $80\,$nm width \cite{Nagel11}, and operation of GBJJ based nanoSQUIDs up to $B=3\,$T \,\cite{Schwarz13} has been demonstrated.
The optimization of the SQUID layouts \cite{Woelbing14} led to the demonstration of ultralow rms flux noise in the thermal white noise limit down to $\sim 45\,{\rm n}\Phi_0/{\rm Hz}^{1/2}$ at $T=4.2\,$K \,\cite{Schwarz15}.
For a MNP placed in 10\,nm-distance to a constriction in the SQUID loop, this corresponds to a spin sensitivity of $\sim 3\,\mu_{\rm B}/{\rm Hz}^{1/2}$ ($\mu_{\rm B}$ is the Bohr magneton).
Such YBCO nanoSQUIDs with bicrystal GBJJs have been used for the investigation of the magnetization reversal of Fe and Co nanowires \cite{Schwarz15, Martinez-Perez18} and Co MNPs \cite{Martinez-Perez17, Martinez-Perez20}, including e.g.~the analysis of MNP switching fields over a wide $T$ range from 300\,mK up to 80\,K.

A significant drawback for high-$T_{\rm c}$ cuprate SQUIDs in general is the very large amount of low-frequceny excess noise, typically scaling with frequency $f$ as $S_\Phi\propto 1/f$ ($1/f$ noise) \cite{Koelle99}.
The dominant source of $1/f$ noise comes from critical current $I_0$ fluctations due to localized defects in the JJ barrier \cite{Rogers84}, which can be several orders of magnitude larger than for conventional Nb tunnel junctions \cite{Foglietti86, Savo87, Marx97}.
For YBCO nanoSQUIDs with ultralow thermal white noise, this excess noise can dominate the spectral density of flux noise up to high frequencies in the MHz range \cite{Schwarz15, Arpaia17}.
An established procedure to suppress this $1/f$ noise contribution from $I_0$ fluctuations in dc SQUIDs is the application of bias reversal schemes (including proper modulation of flux bias), which suppresses $1/f$ noise below the applied bias reversal frequency $f_{\rm br}$ \cite{Koch83, Drung-Mueck-SHB-I.4}.
However, applying such schemes to YBCO nanoSQUIDs is a particular challenge for the SQUID readout electronics and nanoSQUID design, because very large $f_{\rm br}$ is required, and the required modulation of the flux bias points at $f_{\rm br}$ is difficult to achieve due to the small SQUID inductance $L$ and concomitantly small mutual inductance $M$ between a nanoSQUID and the flux modulation line.

We further note that the flux noise spectrum of YBCO nanoSQUIDs often shows a superposition of a few Lorentzians rather than a pure $1/f$ spectrum, which can be explained by the relatively small number of dominating defects in JJs with small size \cite{Schwarz15, Arpaia17}.
In addition we find excess $1/f$ noise in our GBJJ YBCO nanoSQUIDs, which has been attributed to the fluctuations of spins of unknown origin, most likely due to defects at the substrate/YBCO interface or at the edges of the nanopatterned structures \cite{Schwarz15}.
Another open issue is the unclear origin of strong $1/f$ noise in YBCO nanoSQUIDs based on cJJs.
For those devices, this excess noise has been also attributed to $I_0$ fluctuations, although, there is no barrier in such nanowires \cite{Arpaia17}.

Obviously, the strong low-frequency excess noise in YBCO nanoSQUIDs has to be related to the defect structure in the GB barrier or in the YBCO films forming cJJs, together with possible contributions from the YBCO/substrate interface or damaged edges.
Apart from optimizing thin film growth by introducing suitable buffer or cap layers \cite{Ludwig95, Faley06b, Faley17}, a possible solution to this problem may be the growth of multilayers (superlattices), involving epitaxially grown interlayers between YBCO films \cite{Triscone97}, to interrupt the growth of extended defects in YBCO.
Such a multilayer or superlattice approach has been successfully used to significantly enhance the critical current density in YBCO films\cite{Gross90, Foltyn05, Pan06b, Foltyn07}, in particular for thick films developed for high-current applications. 
Accordingly, the focus within this approch was on basic studies and on the improvement of pinning properties of YBCO films, often by deliberately introducing a high density of defects.
Studies on $1/f$ flux noise in YBCO thin films and SQUIDs based on them, also indicated possible improvements via improved pinning properties \cite{Koelle99}.
However, so far, there is no information on the possible modification (improvement or deterioration) of $1/f$ noise from $I_0$ fluctuations in YBCO JJs and SQUIDs based on them, induced by the implementation of superlattice structures into the devices.

In this work, we report on the fabrication and properties of YBCO nanoSQUIDs that are based on a heteroepitaxial YBCO/STO superlattice, grown on a STO bicrystal substrate.
The comparison with single layer YBCO devices of similar geometry shows, that the superlattice nanoSQUIDs yield comparable electric transport properties and comparable upper bounds to the flux noise in the thermal white noise limit.
Regarding low-frequency excess noise, however, we find a strong reduction of the low-frequency noise in the superlattice nanoSQUIDs by more than one order of magnitude in rms flux noise at 1\,Hz.

\section{Device fabrication and layout}
\label{sec:Fab}

We first fabricated very thin $c$-axis oriented epitaxial single layer YBCO thin films by pulsed laser deposition (PLD) on (100)-oriented SrTiO$_3$ (STO) single crystal substrates.
For films of thickness $d_{\rm Y}=7$, 17, 23 and 31\,nm, we find $T_{\rm c}= 61$, 81, 83 and 84\,K, respectively (via inductive $T_{\rm c}$ measurements).
Based on those results, we decided to fabricate multilayer (superlattice) devices with $d_{\rm Y}\approx 30\,$nm per YBCO layer, to ensure a high enough $T_{\rm c}$.

We used PLD to grow an epitaxial YBCO/STO superlattice on a STO bicrystal substrate with a symmetric ($\pm 12^\circ$) [001]-tilt grain boundary (GB), i.e., with a misorientation angle $2\theta = 24^\circ$ ($\theta = 12^\circ$).
The multilayer consists of a stack of four $c$-axis oriented YBCO layers with $d_{\rm Y}\approx 30\,$nm thickness per layer, separated by three STO interlayers with $3\,$nm thickness per layer.
The choice of the total YBCO thickness $d=4d_{\rm Y}\approx 120\,$nm is based on an optimization study, that revealed optimum spin sensitivity for this choice of $d$.\cite{Woelbing14}
For all layers, we used the same deposition parameters (substrate temperature $T_{\rm s}=800\,^\circ$C and oxygen partial pressure $p_{\rm O_2}=0.2\,$mbar).
For details of the growth process of our YBCO thin films on STO and of their structural and electric transport properties see e.g.~Refs.~[\onlinecite{Werner10, Scharinger12, Schwarz13}].
An in-situ evaporated 65-nm-thick Au layer on top of the YBCO/STO superlattice serves as a resistive shunt to provide nonhysteretic current-voltage characteristics (IVCs) of the grain boundary junctions formed in YBCO and as a protection layer during Ga FIB nanopatterning.\cite{Nagel11, Lin20}

\begin{figure}[t]
\includegraphics[width=0.8\columnwidth]{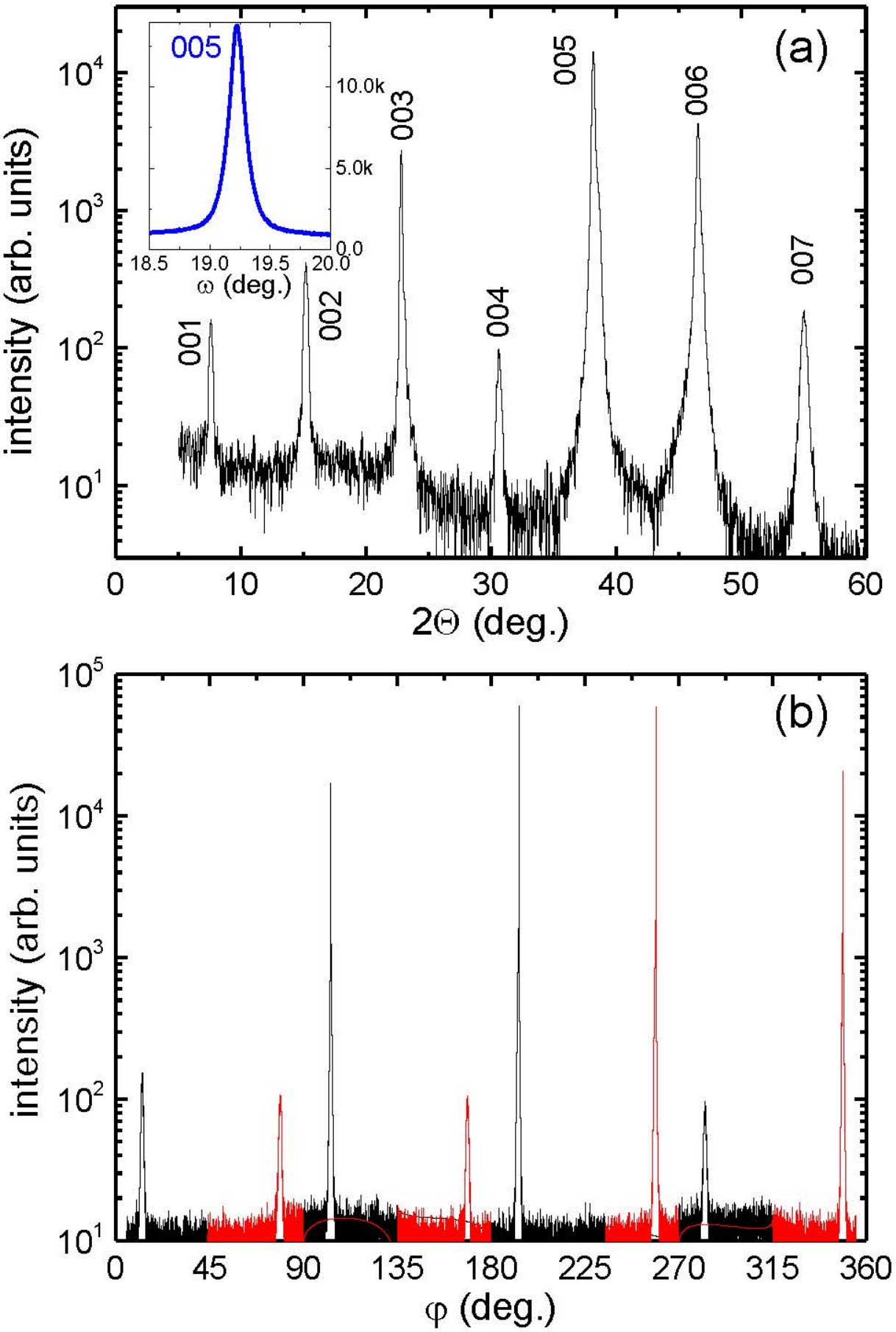}
\caption{XRD data of the unpatterned YBCO/STO superlattice on a STO bicrystal substrate.
(a) $\Theta-2\Theta$ scan, showing YBCO $(00\ell)$ Bragg peaks.
Inset shows rocking curve of the (005) peak.
(b) YBCO (103) plane XRD $\varphi$-scan.
Black and red indicate diffraction peaks from the two grains of the bicrystal.}
\label{Fig:XRD}
\end{figure}

For the characterization of the crystalline quality of the unpatterned YBCO/STO multilayer, we performed x-ray diffraction (XRD) using a 4-circle PhilipsX'Pert diffractometer.
Fig.~\ref{Fig:XRD}(a) shows a $\Theta-2\Theta$ scan, indicating single phase $c$-axis oriented YBCO films.
The $\omega$ scan (rocking curve), shown in the inset of Fig.~\ref{Fig:XRD}(a), yields good alingment of the $c$-axis perpendicular to the substrate plane, with full-width-half-maximum (FWHM) $\sim 0.16^\circ$.
From the position of the YBCO (005) peak we extract a $c$-axis lattice constant of $11.69\,\AA$, which is close to the value of fully oxygenated unstrained YBCO \cite{Jorgensen90}.
Fig.~\ref{Fig:XRD}(b) shows a YBCO (103) $\varphi$-scan with diffraction peaks at $0^\circ$ $\pm$ $(\theta)$, $90^\circ$ $\pm$ $(\theta)$, $180^\circ$ $\pm$ $(\theta)$ and $360^\circ$ $\pm$ $(\theta)$, reflecting the fourfold symmetry of the crystal lattice and rotation of the crystallographic axes by $\pm\theta$ across the GB.
No other peaks can be detected, which confirms in-plane epitaxial growth just according to the orientation of the STO bicrystal without misaligned grains.

After charactierization of the unpatterned YBCO/STO multilayer, we used photolithography and Ar ion milling to prepattern 16 microbridges straddling the grain boundary to create $8\,\mu$m wide GBJJs; those will be referred to as JJ-1 to JJ-16.\cite{Hilgenkamp02}
Those bridges are connected to several-mm long and few-$100\,\mu$m wide contact pads onto which Al wires are bonded as voltage and current leads for 4-point measurements.\cite{Lin20}
Subsequently, YBCO nanoSQUIDs with similar sizes are nanopatterned into some of the $8\,\mu$m-wide GBJJs by FIB milling with 30-keV Ga ions in a dual-beam FIB system.

\begin{figure}[t]
\includegraphics[width=0.8\columnwidth]{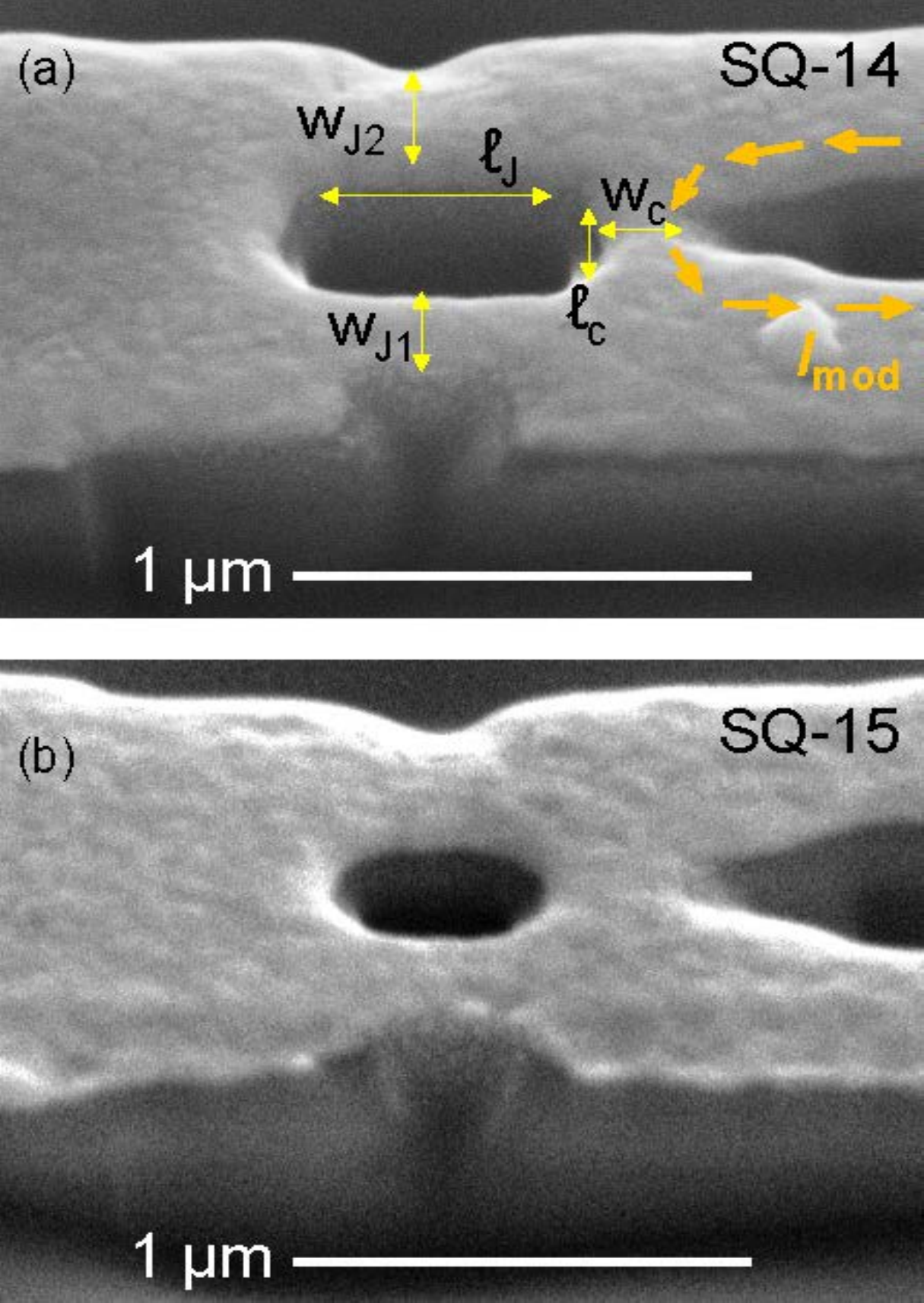}
\caption{SEM images of YBCO/STO superlattice nanoSQUIDs, covered with Au.
The grain boundary (not visible) runs from top to bottom and intersects the SQUID hole to form two JJs ($\sim 250-320\,$ nm wide).
Arrows in (a) indicate the geometric parameters as listed in Table \ref{Tab:samples} and the modulation current $I_{\rm mod}$ that is used to flux-bias the SQUID.}
\label{Fig:SEM}
\end{figure}

Fig.~\ref{Fig:SEM} shows scanning electron microscopy (SEM) images of two YBCO/STO multilayer nanoSQUIDs SQ-14 and SQ-15, nanopatterned into JJ-14 and JJ-15, respectively.
At the position where the GB crosses the microbridge, the Ga FIB is used to define two sub-$\mu$m-wide GBJJs with width $w_{{\rm J}i}$ ($i=1,2$) and to mill the SQUID hole with size $\sim\ell_{\rm J} \times \ell_{\rm c}$.
In addition, we use the Ga FIB to cut a slit perpendicular to the GB (from the right in Fig.~\ref{Fig:SEM}) towards the GB, to produce a constriction with width $w_{\rm c}$ in the SQUID loop.
By applying a modulation current $I_{\rm mod}$ through the constriction (indicated as arrows in Fig.~\ref{Fig:SEM}(a)), the magnetic flux $\Phi$ in the SQUID can be modulated, to ensure SQUID operation at optimum flux bias and to perform SQUID readout in a flux-locked loop (FLL).\cite{Drung-Mueck-SHB-I.4} 
For measurements of magnetization reversal of individual MNPs, placing the MNP on top of the constriction also provides optimum coupling of signal to the nanoSQUID.\cite{Martinez-Perez17a, Martinez-Perez17}

On JJ-15 we determined from resistance vs $T$ measurement $T_{\rm c}=87.2\,$K, with a transition width $\Delta T_{\rm c}\approx 0.8\,$K.
For SQ-15 (nanopatterned into JJ-15), we determined essentially unchanged $T_{\rm c}$ and $\Delta T_{\rm c}$ after FIB patterning.

\section{Transport and noise properties of superlattice JJs and nanoSQUIDs}
\label{sec:JJs-and-nSQ}

In this section, we present and discuss electric transport and noise properties measured in liquid Helium at $T=4.2\,$K in an electrically and magnetically shielded environment.
We focus on two fabricated YBCO/STO superlattice nanoSQUIDs, SQ-14 and SQ-15, which were FIB-patterned from JJ-14 and JJ-15, respectively.
For comparison we also show data for a single layer (SL) 120\,nm-thick YBCO nanoSQUID (SQ-SL), also with a 65\,nm-thick Au layer on top, with comparable lateral geometry.
This SL nanoSQUID was FIB-patterned from the $8\,\mu$m-wide JJ-SL, also on a STO bicrystal substrate with misorientation angle $2\theta=24\,^\circ$.

\begin{figure}[t]
\includegraphics[width=0.8\columnwidth]{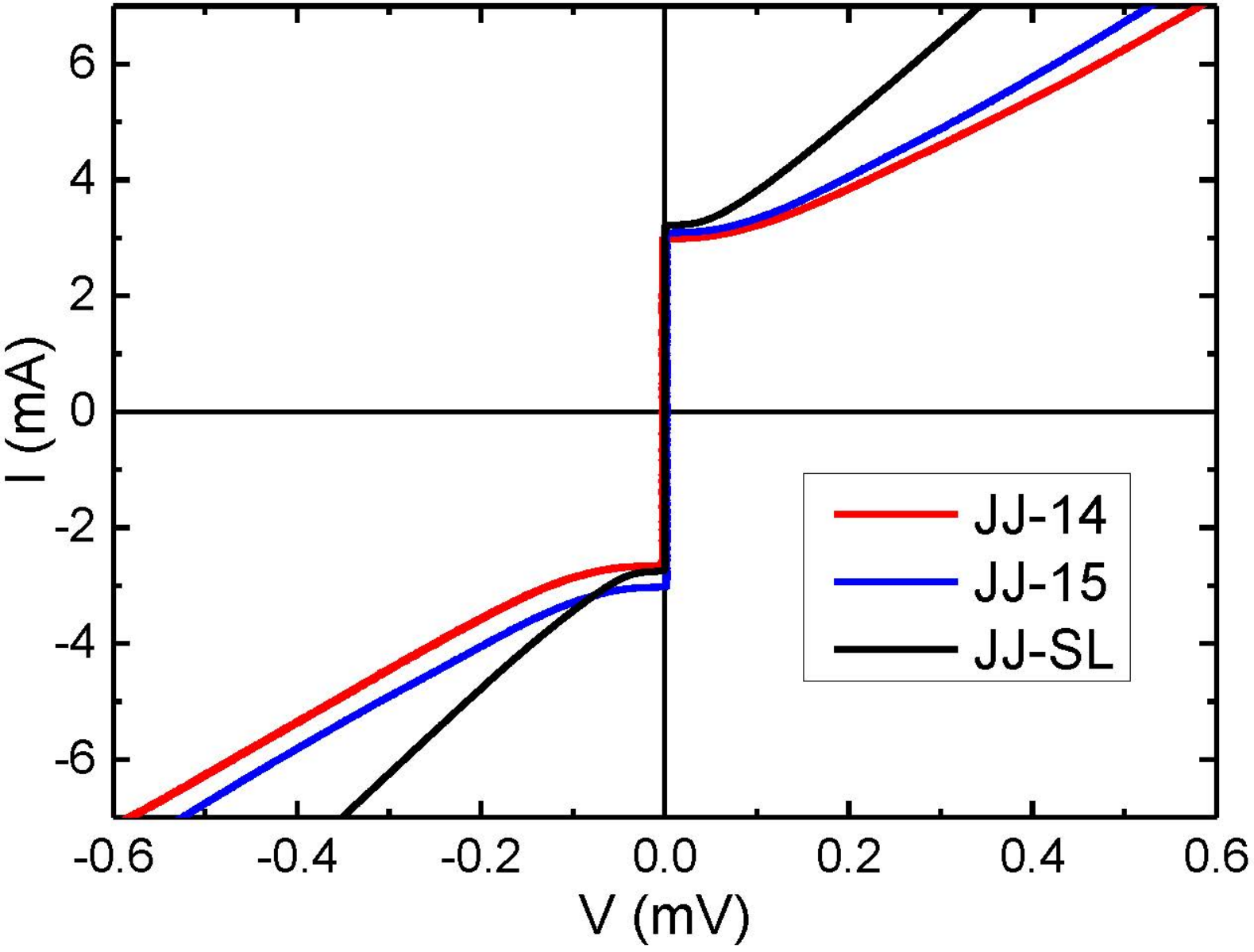}
\caption{IVCs (at $T=4.2\,$K and $H=0$) of 8-$\mu$m-wide GBJJs in a YBCO/STO superlattice (JJ-14 and JJ-15) and in a single layer YBCO film (JJ-SL), all with Au on top.}

\label{Fig:IV-JJs}
\end{figure}

The IVCs of JJ-14 and JJ-15, measured before FIB patterning, are shown in Fig.~\ref{Fig:IV-JJs}.
Those were recorded in zero applied magnetic field ($H=0$).
The IVCs have the typical shape described by the resistively and capacitively shunted junction (RCSJ) model,\cite{Stewart68, McCumber68} without hysteresis.
We note that the $8\,\mu$m-wide JJs are in the long junction limit, as also revealed by $I_{\rm c}(H)$ measurements (not shown).
The $I_{\rm c}(H)$ curves show a slight asymmetry, which means that the maximum critical current $I_{\rm c,max}$ is not reached exactly at $H=0$; this is the reason for the slight asymmetries in the IVCs (different critical current for opposite polarity).
Hence, the $I_{\rm c,max}$ values quoted in Table \ref{Tab:samples} have been obtained from the maxima of the $I_{\rm c}(H)$ patterns.
Values of $I_{\rm c,max}$ for both JJs are around 3\,mA.
Converting this into a critical current density $j_{\rm c}$, by dividing by the JJ width ($8\,\mu$m) and total YBCO film tickness (120\,nm) yields $j_{\rm c}\approx 3\times 10^5\,{\rm A/cm}^2$ (see also Table \ref{Tab:samples}).
This is a typical value which we obtain for YBCO GBJJs of similar size, made from single layer films\cite{Nagel11, Lin20} and hence indicates, that the current flowing through the JJs is distributed across all four YBCO layers in the superlattice.
This observation is also consistent with microstructural studies on YBCO/STO superlattices with 3\,nm-thick STO layers, which showed discontinuities in the very thin STO layers \cite{Ryen98}.

\renewcommand{\arraystretch}{1.1}
\begin{table*}[t]
\caption{Summary of geometric and electric parameters of $8\,\mu$-wide JJs and nanoSQUIDs as defined in the text. 
JJ-14 and JJ-15 are based on YBCO/STO superlattices, from which SQ-14 and SQ-15 was nanopatterned, respectively.
JJ-SL is based on a single layer of  YBCO, from which SQ-SL was nanopatterned.
For all devices the total YBCO thickness is 120\,nm and the Au layer on top is 65\,nm thick.}
\begin{center}
\tabcolsep1.5mm
\begin{tabular}{c c c c c c c c c c c c c c c}\hline\hline
device	&$w_{\rm J1}$	&$w_{\rm J2}$	&$w_{\rm c}$	&$\ell_{\rm c}$	&$\ell_{\rm J}$	&$I_{\rm c,max}$	&$R_{\rm n}$	&$V_{\rm c}$	&$j_{\rm c}$			&$I_{\rm c,min}$	&$I_{\rm mod,0}$	&$M$	&$L$	&$\beta_L$	\\
		&(nm)        		&(nm)			&(nm)		&(nm)			&(nm)			&(mA)			&($\Omega$)	&(mV)		&($10^5$\,A/cm$^2$)	&(mA)			&(mA)			&(pH)	&(pH)	& 	\\\hline\hline
JJ-14	& 				&       			&     		& 				& 				& 3.2			& 0.091		& 0.29		& 3.3				&				&				&		&		&	\\\hline
JJ-15	& 				&       			&    			& 				& 				& 3.3 			& 0.098 		& 0.32		& 3.4				&				&				&		&		&	\\\hline                                                                                                                                                         
JJ-SL 	&     			&       			&     		& 				& 				& 3.4			& 0.07		& 0.23		& 3.5				&				&				&		&		&	\\\hline\hline                                                                                                                                                  
SQ-14	& 280     		& 320     		& 180    		& 250			& 500			& 0.42			& 0.82		& 0.34		& 5.9				& 0.26			& 0.90	   		& 2.3	& 7.1	& 1.5\\\hline
SQ-15	& 250    			& 280      		& 280    		& 200			& 400			& 0.29			& 1.28		& 0.37		& 4.5				& 0.12			& 2.19	    		& 1.0	& 4.9	& 0.7\\\hline                                                                                                                           
SQ-SL 	& 280			& 290      		& 300		& 350			& 400			& 0.20			& 1.68		& 0.34		& 3.0				& 0.12			& 1.24			& 1.7	& 12		& 1.2\\\hline
\end{tabular}
\end{center}
\label{Tab:samples}
\end{table*}

With a measured resistance $R_{\rm n}\sim 0.1\,\Omega$ (which is dominated by the shunting Au layer \cite{Lin20}) for the two JJs, we obtain a characteristic voltage $V_{\rm c}=I_{\rm c,max}R_{\rm n}\sim 0.3\,$mV (see also Table \ref{Tab:samples}); this is at the lower end of the range of values of $V_{\rm c}$ that we observe for single layer YBCO nanoSQUIDs based on GBJJs on STO bicrystal substrates.\cite{Woelbing14}
For comparison, the IVC of JJ-SL is also shown in Fig.~\ref{Fig:IV-JJs}; this device indeed has similar values for $I_{\rm c}$, $R_{\rm n}$ and $V_{\rm c}$; c.f.~Table \ref{Tab:samples}.
We note that values of $I_{\rm c,max}$ and $R_{\rm n}$ for the other $8\,\mu$m-wide JJs on the same chip with the YBCO/STO superlattice are very similar to JJ-14 and JJ-15.

\begin{figure}[t]
\includegraphics[width=0.8\columnwidth]{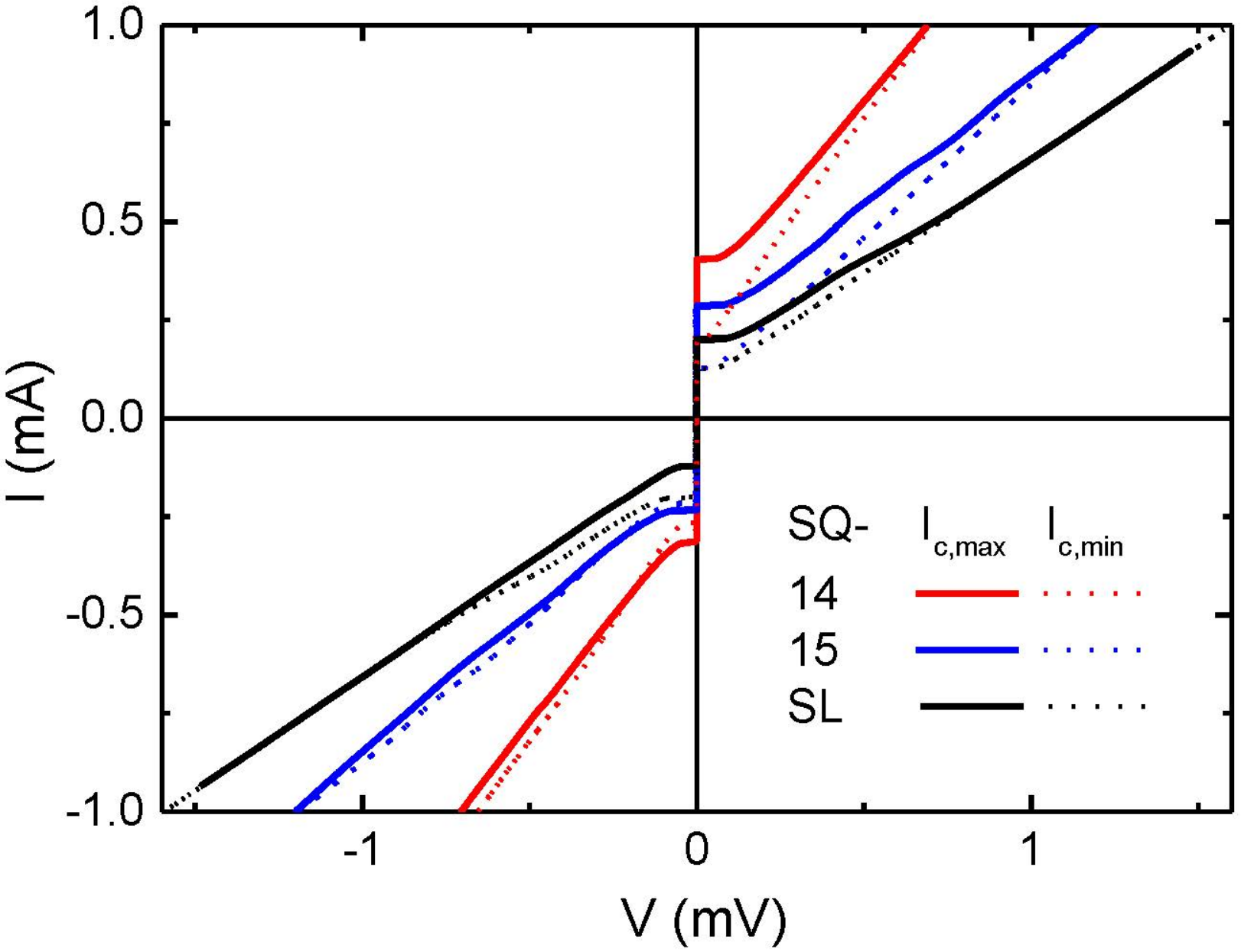}
\caption{IVCs of superlattice SQUIDs (SQ-14 and SQ-15) and single layer SQUID (SQ-SL).
Solid (dashed) lines are recorded with $I_{\rm mod}$ adjusted to obtain maximum (minimum) critical current $I_{\rm c,max}$ for $I>0$.}
\label{Fig:IV-SQs}
\end{figure}

Fig.~\ref{Fig:IV-SQs} shows IVCs of superlattice nanoSQUIDs SQ-14 and SQ-15 and, for comparison, the IVC of the single layer nanoSQUID SQ-SL with comparable geometry.
For those measurements, $I_{\rm mod}$ has been adjusted to yield maximum critical current $I_{\rm c,max}$ (solid lines) and minimum critical current $I_{\rm c,min}$ (dashed lines) on the positive branches.

Geometric and electrical parameters for all three SQUIDs are summarized in Table \ref{Tab:samples}, together with parameters from the JJs obtained before nanoSQUID patterning.
Also after FIB nanopatterning, we find nonhysteretic RCSJ-type IVCs, with values for $I_{\rm c,max}$, $R_{\rm n}$ and $V_{\rm c}$ which are comparable to GBJJ nanoSQUIDs from single layer YBCO films with similar geometry.\cite{Lin20}
We note that we observe for all nanoSQUIDs slightly larger $j_{\rm c}$ and $V_{\rm c}$ values, as compared to those obtained from the $8\,\mu$m-wide JJs.
This is typical for all our YBCO nanoSQUIDs (see e.g.~Ref.~\onlinecite{Nagel11}), and we attribute this to the fact, that the inhomogeneity of the GB (e.g.~due to faceting \cite{Hilgenkamp02}) is slightly reduced upon reducing the JJ width to the deep sub-$\mu$m regime.

\begin{figure}[b]
\includegraphics[width=0.8\columnwidth]{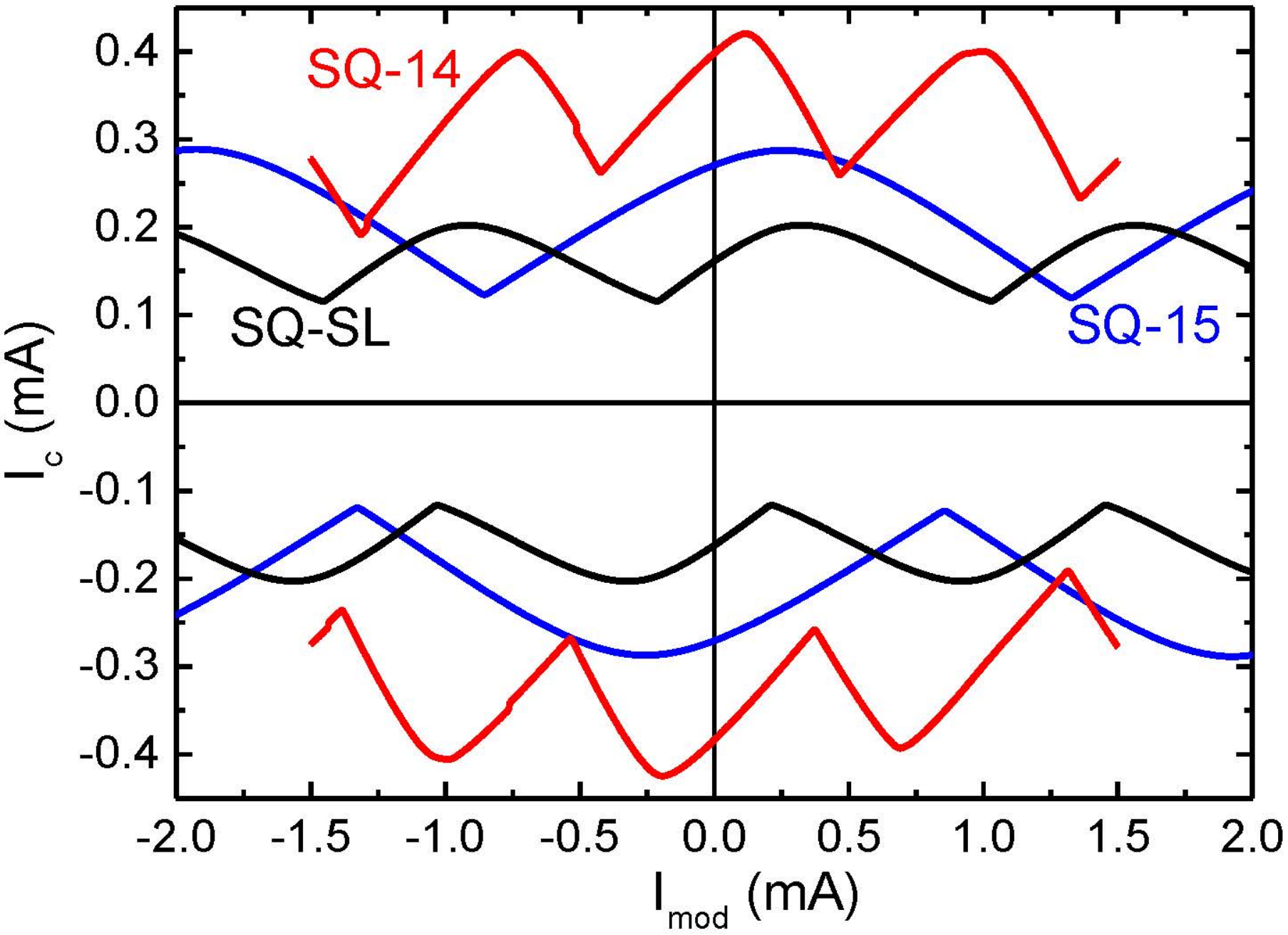}
\caption{Critical current $I_{\rm c}$ vs.~$I_{\rm mod}$ of multilayer SQUIDs (SQ-14 and SQ-15) and single layer SQUID (SQ-SL) for both bias current polarities.}
\label{Fig:Ic-Imod}
\end{figure}

Fig.~\ref{Fig:Ic-Imod} shows critical current $I_{\rm c}$ vs $I_{\rm mod}$ oscillations of all three nanoSQUIDs. 
From the modulation period $I_{\rm mod,0}$ we determine the mutual inductance $M=\Phi_0/ I_{\rm mod,0}$.
Obviously, the mutual inductance for SQ-14 is more than a factor of two larger than for SQ-15, while SQ-SL has a value of $M$ inbetween.
This observation is consistent with the different values for the constriction width $w_{\rm c}$ (cf.~Table \ref{Tab:samples} and Fig.~\ref{Fig:SEM}).
We find that a narrower constriction yields a larger $M$.
This is also supported by inductance calculations based on simulations of the supercurrrent density distribution in our nanoSQUIDs via the software 3D-MLSI, which solves the London equation in 2-dimensional current sheets \cite{Khapaev03,Woelbing14,Nagel11a}.
A larger $M$ is beneficial, as it relaxes requirements (maximum feedback and modulation currents) on the SQUID readout electronics for FLL operation and as it also improves the coupling between a MNP and a nanoSQUID, resulting in improved spin sensitivity.

To obtain a rough estimate of the inductance $L$ of the nanoSQUIDs, we determine the screening parameter $\beta_L\equiv 2LI_0/\Phi_0$ from the modulation depth $\Delta I_{\rm c}\equiv I_{\rm c,max}-I_{\rm c,min}$ of the  $I_{\rm c}(I_{\rm mod})$ oscillations.
Into the definition of $\beta_L$ enters the noise free critical current $I_0$ of the JJs.
In the case of negligible noise rounding, as applicable to our devices at 4.2\,K (cf.~Fig.~\ref{Fig:IV-SQs}), we can replace $2I_0$ by $I_{\rm c,max}$.
Then, by using the dependence $\Delta I_{\rm c}/I_{\rm c,max}(\beta_L)$ derived from numerical simulations for symmetric dc SQUIDs in the noise-free case, we determine $\beta_L$ values as listed in Table \ref{Tab:samples} and which are close to the value $\beta_L\approx 1$ for optimum flux noise\cite{Tesche77, Chesca-SHB-I.2}.
From the estimated values for $\beta_L$ and the measured values for $I_{\rm c,max}$ we then obtain the values for $L$ as listed in Table \ref{Tab:samples}.
For the superlattice nanoSQUIDs we obtain values of $L$ below 10\,pH.
Achieving such a small inductance is important for obtaining very low values of flux noise for the nanoSQUIDs in the thermal white noise limit well below $S_\Phi^{1/2}=1\,\mu\Phi_0/{\rm Hz}^{1/2}$.

We note that all $I_{\rm c}(I_{\rm mod})$ curves in Fig.~\ref{Fig:Ic-Imod} show a clear asymmetry.
This is visible as a shift of the patterns along the $I_{\rm mod}$ axis in opposite direction for opposite current polarities and a skewness of the $I_{\rm c}(I_{\rm mod})$ curves, which can arise from asymmetries in the critical currents of the two JJs and from inductance asymmetry \cite{Tesche77, Chesca-SHB-I.2}.
These observations are consistent with the slight asymmetry in the widths of the JJs, inducing a critical current asymmetry, and with the fact that the constrictions in the SQUIDs induce an inductance asymmetry.

\begin{figure}[b]
\includegraphics[width=0.8\columnwidth]{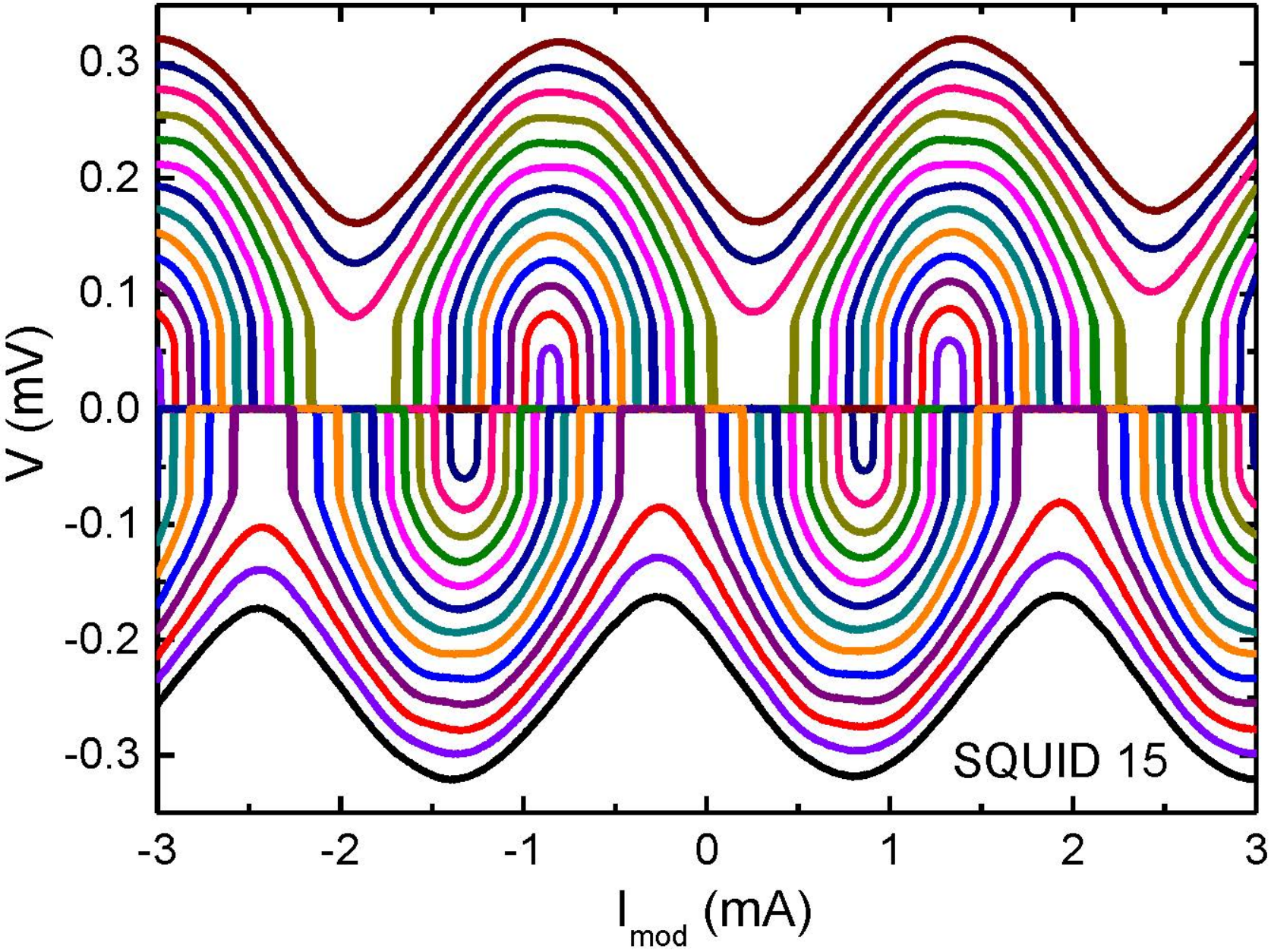}
\caption{Voltage-flux characteristics $V(I_{\rm mod})$ of SQ-15 at different bias currents $I_{\rm b}$ from 133 to 360\,$\mu$A in $\sim 19\,\mu$A steps (for both polarities).}
\label{Fig:V-Imod}
\end{figure}

Fig.~\ref{Fig:V-Imod} shows V($I_{\rm mod}$) oscillations of SQ-15, measured at different bias currents.
The shift of those curves along the $I_{mod}$ axis for opposite bias current polarity is consistent with the concomitant shift in the $I_{\rm c}(I_{\rm mod})$ pattern for SQ-15 shown in Fig.~\ref{Fig:Ic-Imod}.
Moreover, the maxima in V($I_{\rm mod}$) are slightly shifting along the $I_{mod}$ axis with increasing bias current, which is consistent with a small inductance asymmetry, as discussed above.
We only show here V($I_{\rm mod}$) oscillations for SQ-15; however, the same features are also present for SQ-14 and SQ-SL.

\begin{figure}[t]
\includegraphics[width=0.8\columnwidth]{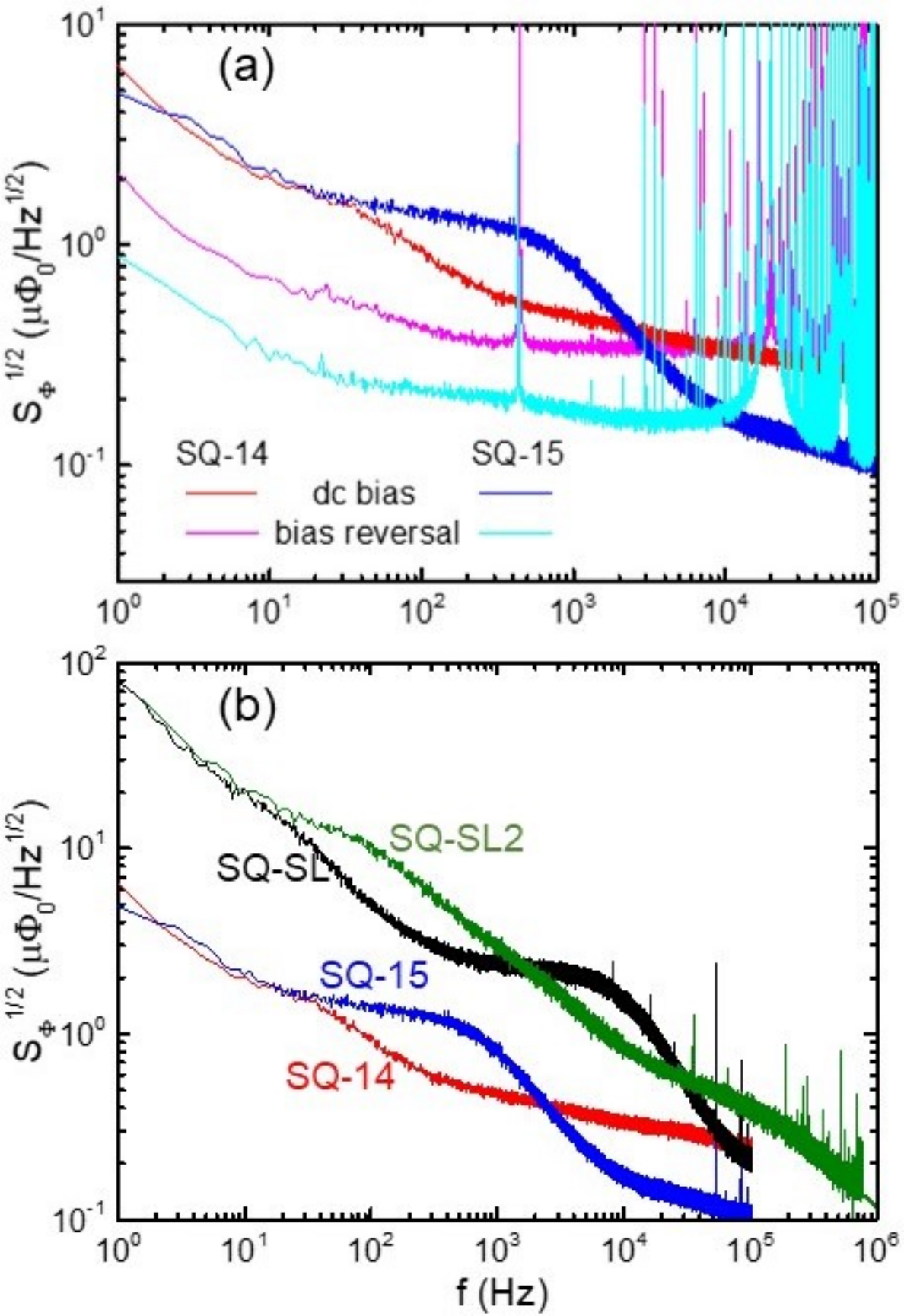}
\caption{Rms spectral density of flux noise $S_\Phi^{1/2}(f)$ measured in FLL mode: (a) data for SQ-14 and SQ-15, measured with dc bias and bias reversal ($f_{\rm br}=20\,$kHz).
(b) dc bias readout data of SQ-14, SQ-15, together with SQ-SL and SQ-SL2 on an expanded scale for $S_\Phi^{1/2}$ and $f$.}
\label{Fig:noise}
\end{figure}

Finally, we discuss the flux noise of the superlattice YBCO nanoSQUIDs, which has been measured in FLL mode.
Fig.~\ref{Fig:noise}(a) shows the rms spectral density of flux noise $S_\Phi^{1/2}(f)$ of SQ14 and SQ-15 measured with dc bias and with bias reversal at frequency $f_{\rm br}=20\,$kHz.
In the bias reversal mode, the contribution of critical current fluctuations to low-frequency excess noise is removed below $f_{\rm br}$.\cite{Drung-Mueck-SHB-I.4}
We note that we do not reach the thermal white noise regime even at the highest frequency of 100\,kHz up to which we performed measurements.
Hence the rms flux noise values $S_\Phi^{1/2}(100\,{\rm kHz})=244\,{\rm n}\Phi_0/{\rm Hz}^{1/2}$ for SQ-14 and $104\,{\rm n}\Phi_0/{\rm Hz}^{1/2}$ for SQ-15 (with dc bias readout) are upper bounds for the thermal white noise limit.
Those are comparable to the best values for the flux noise at high frequencies obtained for YBCO nanoSQUIDs based on GBJJs in single layer devices.\cite{Schwarz15, Lin20}

Most importantly, with dc bias readout we obtain at $f= 1\,$Hz values $S_\Phi^{1/2}(1\,{\rm Hz})=6.5\,\mu\Phi_0/{\rm Hz}^{1/2}$ for SQ-14 and $4.9\,\mu\Phi_0/{\rm Hz}^{1/2}$ for SQ-15.
Those values are more than an order of magnitude lower than what we obtained so far for single layer YBCO nanoSQUIDs on STO bicrystal substrates \cite{Schwarz15}, and they are comparable to the values for single layer YBCO nanoSQUIDs on MgO bicrystal substrates, as reported very recently.\cite{Lin20}
To illustrate this observation, we show in \ref{Fig:noise}(b) again the rms flux noise measured in dc bias mode for SQ-14 and SQ-15, now together with noise data for the single layer devices SQ-SL and SQ-SL2.
The latter has also been fabricated on an STO bicrystal substrate with an Au layer on top.
SQ-SL2 has been reported earlier to show the so far lowest flux noise of $\sim 45\,{\rm n}\Phi_0/{\rm Hz}^{1/2}$ in the thermal white noise limit (at very high frequency $>7\,$MHz) for any of our YBCO nanoSQUIDs \cite{Schwarz15}.
The flux noise at $f=1\,$Hz for both single layer devices shown in Fig.~\ref{Fig:noise}(b) is $\sim80\,\mu\Phi_0/{\rm Hz}^{1/2}$.
This observation shows, that the use of YBCO/STO superlattices instead of YBCO single layer films can significantly reduce $1/f$ noise in YBCO nanoSQUIDs based on GBJJs.

The noise spectra for SQ-14 and SQ-15 in bias reversal mode yield a significant improvement of $S_\Phi^{1/2}(1\,{\rm Hz})$ over dc bias readout, for SQ-15 even below  $1\,\mu\Phi_0/{\rm Hz}^{1/2}$ [cf.~Fig.~\ref{Fig:noise}(a)].
This shows, that $I_0$ fluctutations in the GBJJ barriers are the major source of low-frequency excess noise in our devices, stemming from defects in the barriers.
Hence, we conclude that the significantly reduced low-frequency excess noise in the YBCO/STO superlattice nanoSQUIDs is most likely due to an improved quality of the grain boundary, as compared to single layer YBCO nanoSQUIDs.
Finally, we note that we still see low-frequency excess noise even with bias reversal readout.
This issue has been addressed already in Ref.~\onlinecite{Schwarz15}, where we attributed this to possible contributions from fluctuating spins in the substrate close to the STO/YBCO interface.

\section{Conclusions}
\label{sec:Conclusions}

We have fabricated YBCO dc nanoSQUIDs from a YBCO/STO superlattice, consisting of four 30\,nm-thick individual YBCO layers, separated by 3\,nm-thick STO layers.
The superlattice is grown heteroepitaxially on a STO bicrystal substrate with $24\,^\circ$ misorientation angle, and covered with 65\,nm-thick Au on top, as a resistive shunt and for protection during Ga FIB milling.
The characterization of crystalline film quality and measurement of the electric transport properties before and after Ga FIB nanopatterning shows that the superlattice devices have comparable quality as for single layer devices with the same total YBCO fillm thickness.
Also the measured noise properties in the thermal withe noise limit are similar for single layer and superlattice devices.
This is in strong constrast to the observed low-frequency excess noise: superlattice devices yield more than one order of magnitude lower noise at 1\,Hz as compared to single layer devices.

Because the low-frequency excess noise is dominated by fluctuations of the critical current $I_0$ in the grain boundary Josephson junctions, we attribute the improved low-frequency noise performance of the superlattice nanoSQUIDs to an improved microstructure of the grain boundaries forming the Josephson junctions.
However, so far we have no direct information on the microstructure at the grain boundaries in our superlattices.
One possible reason for the improvement might be due to reduced facetting \cite{Hilgenkamp02} of the grain boundaries.
Certainly, a clarification of this issue, or of other possible modifications in the defect structure of the grain boundaries, that are induced by inserting STO interlayers in the YBCO films, is required in future studies.

In any case, the achieved improvement in low-frequency excess noise in YBCO nanoSQUIDs provides the opportunity to realize ultrasensitive devices for scanning SQUID microscopy and for the investigation of magnetization reversal processes in individual magnetic nanosystems, by utilizing already achievable ultralow levels of white noise and expand those on the frequency scale down to well below the MHz range.
Moreover, the superlattice approach may also be helpful to improve the low-frequency noise performance of other devices, e.g.~sensitive SQUID magnetometers, that are based on grain boundaries.
It remains to be shown, whether this superlattice approach could also improve the low-frequency excess noise in YBCO nanoSQUIDs based on nanowires (constriction junctions).

\section*{Conflicts of interest}

The authors declare no conflicts of interest

\acknowledgments

J.~Lin acknowledges funding by the China Scholarship Council (CSC). 
We gratefully acknowledge fruitful discussions and technical support by M.~Turad and R.~L\"offler (LISA$^+$) and by C.~Back.
This work was supported by the  COST action NANOCOHYBRI (CA16218).

\bibliography{YBCO-STO-nSQUID}

\end{document}